# Logistic Regression Modeling Based on Fractal Dimension Curves of Urban Growth

Yanguang Chen

(Department of Geography, College of Urban and Environmental Sciences, Peking University, 100871, Beijing, China. Email: chenyg@pku.edu.cn)

**Abstract:** Fractal dimension is an effective scaling exponent of characterizing scale-free phenomena such as cities. Urban growth can be described with time series of fractal dimension of urban form. However, how to explain the factors behind fractal dimension sequences that affect fractal urban growth remains a problem. This paper is devoted to developing a method of logistic regression modeling, which can be employed to find the influencing factors of urban growth and rank them in terms of importance. The logistic regression model comprises three components. The first is a linear function indicating the relationship between time dummy and influencing variables. The second is a logistic function linking fractal dimension and time dummy. The third is a ratio function representing normalized fractal dimension. The core composition is the logistic function that implies the dynamics of spatial replacement. The logistic regression modeling can be extended to other spatial replacement phenomena such as urbanization, traffic network development, and technology innovation diffusion. This study contributes to the development of quantitative analysis tools based on the combination of fractal geometry and conventional mathematical methods.

## 1 Introduction

There are more than one factor that affects urban growth, and there are also various methods to reveal the influencing factors of urban growth. Qualitative analysis, simple statistical analysis, and mathematical modeling analysis can be used to identify the factors affecting city development. Qualitative analysis cannot determine the significance of influencing factors, and it is also difficult to prioritize the influencing factors. To solve this problem, scholars make use of multiple linear regression analysis, especially stepwise regression analysis. To this end, it is necessary to define a measure that reflects urban growth. Urban population, wealth, and urbanized area become the primary indicators (Arbesman, 2012; Dendrinos, 1992; Nordbeck, 1971Woldenberg, 1973). However, due to scale-free property of urban form (Batty and Longley, 1994; Frankhauser, 1994), urban size and area cannot be objectively determined, say nothing about urban wealth (Chen and Lin, 2009). In this case, fractal dimension can be utilized to measure the space filling degree of cities. The time series of fractal dimension of urban form in different years take on sigmoid curve. Thus logistic function can be employed to model urban growth (Chen, 2012). The prediction model of urban growth based on fractal dimension emerge (Chen, 2018). In this type of models, the independent variable is time, and the dependent variable is fractal dimension of urban form.

If there is no real causal relationship between an independent variable and a dependent variable, then the independent variable belongs to a dummy variable. Dummy variables include time variable, distance variable, and categorical variables (nominal variable, indicator variables). Time variable is termed time dummy (Diebold, 2007). The time dummy variable usually suggests real influencing factors of urban growth underlying the temporal variable. These influencing factors do not form a direct linear relationship with the fractal dimension sequence, and simple multiple linear regression analysis cannot be used to determine the importance, priority, and order of the influencing factors. To solve this problem, this paper is devoted to deriving a logistic regression model based on fractal dimension series and multiple explanation variables. The research goal is to construct a framework for logistic regression analysis for geographical systems. In Section 2, two sets of logistic regression models are introduced into urban study by mathematical derivation and analogical analysis. In Section 3, the framework of logistic regression model is outlined, the method is extended to urbanization and transport network research, and several related questions are discussed. Finally, in Section 4, the discussion is concluded by summarizing the main points of this work.

## 2 Models

### 2.1 Logistic regression modeling based on fractal dimension curve

Due to the scaling invariance nature of urban structure, conventional measures such as perimeter, area, density, etc., cannot effectively describe urban morphology. Fractal dimension can be employed to characterize degrees of space filling, spatial evenness, and spatial dependence. Spatial evenness and spatial difference are two sides of the same coin. A sample path of a time series of fractal dimension can be gained by means of observational data of urban form at different times. Because of squashing effect, a sample path of fractal dimension sequence forms an S-liked curve, which is termed fractal dimension curve of urban growth (Figure 1). The fractal dimension curve can be modeled by a type of sigmoid function (Chen, 2012; Chen, 2018). The simplest and most common sigmoid function is the logistic function (Mitchell, 1997). Therefore, in many case, a fractal dimension curve of urban form and growth can be expressed as

$$D(t) = \frac{D_{max}}{1+(D_{max}/D_{(0)}-1)e^{-kt}}, \quad (1)$$

in which $D(t)$ refers to the fractal dimension of urban form at time $t$, $D_{(0)}$ is the initial value of fractal dimension at time $t=0$, $D_{max}$ is the capacity parameter of fractal dimension, i.e., the upper limit of fractal dimension, and $k$ is the initial growth rate of fractal dimension.

The logistic model of fractal dimension curve of urban form based on time series can be used to predict urban growth. By using equation (1), it is possible to estimate the carrying capacity of urban fractal dimension and determine the peak of urban growth rate. Generally speaking, when $D(t)=D_{max}/2$, urban growth rate reaches its peak. Moreover, the urban growth process can be divided into four stages by means of equation (1). However, using equation (1) we cannot explain urban growth rate and different stages. In equation (1), the variable of time, $t$, is a dummy variable, which is termed time dummy (Diebold, 2007). A dummy variable is usually not a true explanatory variable, but a substitute for the explanatory variables. The true explanatory variables may be hidden behind the dummy variable. Suppose that there are $m$ real explanatory variables behind time dummy, that is, $x_j$ ($j=1, 2,…, m$). To reveal the true explanatory variables, let's consider the simplest case where the dummy variable is a linear function of set of explanatory variables. Thus we have a linear decomposition relation as below:

$$t = \frac{1}{k}(c + b_1 x_1(t) + b_2 x_2(t) + \cdots + b_m x_m(t)), \quad (2)$$

where $m$ denotes the number of influence factors, $c$ is a constant, and $b_j$ is the $j$th linear regressive coefficient. Substituting equation (2) into equation (1) yields a logistic regression model based on fractal dimension as below

$$\frac{D(t)}{D_{\max}} = \frac{1}{1+e^{-a-b_1 x_1(t)-b_2 x_2(t)-\cdots-b_m x_m(t)}} = \frac{1}{1+\exp(-\sum_{j=0}^{m} b_j x_j(t))}, \quad (3)$$

in which $a$ and $b_j$ refers to logistic regression coefficients, $a=b_0$, $x_0\equiv 1$ ($j=0$), and specially normalized fractal dimension $D(t)/D_{\max}$ represents fractal dimension ratio, indicating space-filling ratio and spatial evenness degree. Fractal dimension ratio can be expressed as $Q(t)= D(t)/D_{\max}$. The parameter $a$ can be expressed as

$$a = b_0 = c - \ln(\frac{D_{\max}}{D_0}/-1). \quad (4)$$

Using the symbol $b_0$ is only for simple expression. Accordingly, fractal redundancy can be defined as

$$1 - \frac{D(t)}{D_{\max}} = \frac{\exp(-\sum_{j=0}^{m} b_j x_j(t))}{1+\exp(-\sum_{j=0}^{m} b_j x_j(t))}, \quad (5)$$

where fractal redundancy, $1- D(t)/D_{\max}$, implies space-saving ratio and spatial difference degree. Based on equations (3) and (5), a logit transformation of fractal dimension odds can be obtained as follows

$$\ln \frac{D(t)}{D_{\max} - D(t)} = \ln O(t) = a + \sum_{j=1}^{m} b_j x_j(t), \quad (6)$$

where $D(t)/(D_{\max} -D(t))$ represent fractal dimension odds. It can be expressed as

$$O(t) = \frac{D(t)}{D_{\max} - D(t)}, \quad (7)$$

in which $O(t)$ denotes fractal dimension odds or space-saving ratio of time $t$. For simplicity, let $D_{\max} =d$, where $d$ denotes the Euclidean dimension of embedding space. Thus, equations (6) changes to the following form

$$\ln \frac{D(t)}{d - D(t)} = \sum_{j=0}^{m} b_j x_j(t) = a + \sum_{j=1}^{m} b_j x_j(t). \quad (8)$$

Generally speaking, $d=2$. Using equation (6) or (8), we can make a logistic regression analysis based on time series of fractal parameters and related observed data of possible explanatory variables.

Logistic modeling of fractal dimension curves based on time series can be generalized to cross-sectional data. A cross-sectional dataset in a geographical region can be obtained according to city rank at given time. In theory, city size distribution in an urban system corresponds to urban growth process. Small cities represent youth cities, while large cities represent adult cities (Batty and Longley, 1994). Based on rank-size distribution, equation (5) can be revised as

$$\ln \frac{D(r)}{D_{\max} - D(r)} = a + \sum_{j=1}^{m} b_j x_j(r). \qquad (9)$$

where $r$ denotes city rank, $D(r)$ is the fractal dimension of urban form of the city of rank $r$. Accordingly, equation (6) can be rewritten as

$$\ln \frac{D(r)}{2 - D(r)} = \ln O(r) = a + \sum_{j=1}^{m} b_j x_j(r). \qquad (10)$$

Using equation (9) or (10), we can make a logistic regression analysis based on rank series of fractal parameters and related observed data. By means of regression results, it is possible to determine which factors significantly affect urban growth and which factors do not have a significant impact on urban growth. Among the factors that have a significant impact, the primary and secondary factors can be identified and ranked. So, it becomes possible to explain the process of urban growth.

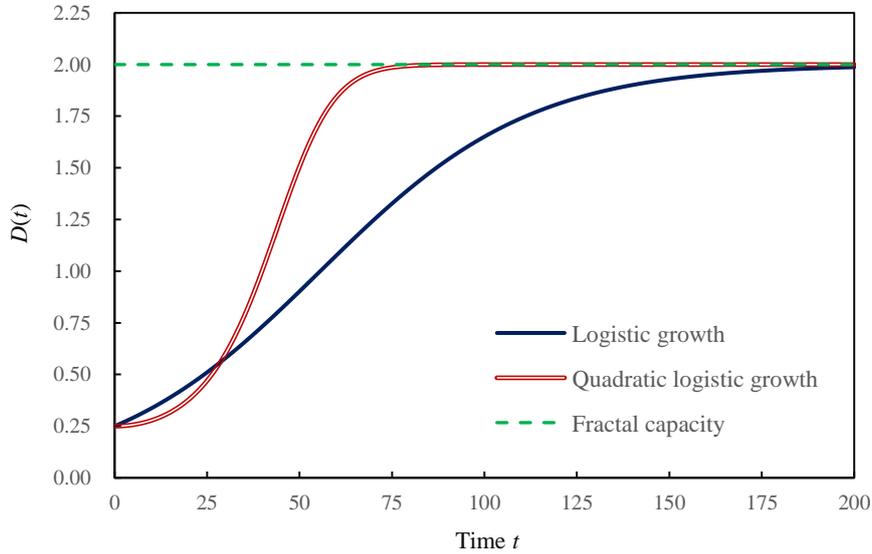

**Figure 1 A diagrammatic sketch for sigmoid growth and squashing effect of fractal dimension**

**Note**: The parameter values of both logistic model and quadratic logistic model of fractal dimension growth are as follows, $D_{\max}=2$, $D_0=0.25$, $k=0.035$. The lower limit of fractal dimension is the topological dimension, $d_T=0$, and the upper limit is the Euclidean dimension of embedding space, $d_E=2$. The squashing of topological dimension and Euclidean dimension make fractal dimension increase along an S-liked curve.

## 2.2 Logistic regression modeling based on quadratic fractal dimension curve

A mathematical model has its effective scope of application. There is no absolutely general models for social and economic phenomena. The conventional logistic function can be used to describe the fractal dimension curves of European and American cities, as well as the fractal dimension curves of some cities along the southeast coast of China. However, this model is not suitable for most mainland Chinese cities. It is not suitable for the majority of Chinese cities, especially northern Chinese cities. The fractal dimension curves of great majority of Chinese cities can be modeled by quadratic logistic function as follows (Chen, 2018)

$$D(t) = \frac{D_{max}}{1+(D_{max}/D_0-1)e^{-(kt)^2}}, \quad (11)$$

which is similar in macro structure to equation (1). If square of time, $t^2$, is a linear function of sets of dependent variables, we have

$$t^2 = \frac{1}{k^2}(c+b_1 x_1(t)+b_2 x_2(t)+\cdots+b_m x_m(t)). \quad (12)$$

Substituting equation (12) into equation (11) yields equation (3). The corresponding logistic regressive model is the same as equation (6), which can be replaced by equation (8) for simplicity.

The mathematical expressions of logistic regression models for quadratic logistic growth remain to be determined by further research. Another possibility is that time, $t$, rather than square of time, $t^2$, is a linear function of multiple explanatory variables, that is, the relation between time dummy and arguments can be expressed by equation (2) instead of equation (12). Substituting equation (2) into equation (11) yields

$$\sqrt{\ln(\frac{D(t)}{D_{max}-D(t)}/\frac{D_0}{D_{max}-D_0})} = c + \sum_{j=1}^{m} b_j x_j(t). \quad (13)$$

For simplicity, equation (13) can be replaced by

$$\sqrt{\ln(\frac{D(t)}{d-D(t)}/\frac{D_0}{d-D_0})} = \sqrt{\ln(\frac{O(t)}{O_0})} = c + \sum_{j=1}^{m} b_j x_j(t). \quad (14)$$

Only through statistic experiments based on observation data can we determine whether to use equations (6) and (8) or equations (13) and (14). The time series data can be replaced by cross-sectional data, thus equation (13) can be re-expressed as

$$\sqrt{\ln(\frac{D(r)}{D_{\max}-D(r)}/\frac{D_0}{D_{\max}-D_0})}=c+\sum_{j=1}^{m}b_j x_j(r). \tag{15}$$

For simplicity, equation (15) can be changed to the following form

$$\sqrt{\ln(\frac{D(r)}{d-D(r)}/\frac{D_0}{d-D_0})}=\sqrt{\ln(\frac{O(r)}{O_0})}=c+\sum_{j=1}^{m}b_j x_j(r). \tag{16}$$

Equation (15) and (16) can be applied to fractal dimension dataset of an urban system to make horizontal logistic regression analysis.

## 3 Discussion

The basic framework of logistic modeling for fractal dimension curves have been outlined above. The methodological framework comprises three components, which can be abstracted as three mathematical equations. The first component is a linear equation, that is

$$z=b_0+b_1 x_1(t)+b_2 x_2(t)+\cdots+b_m x_m(t). \tag{17}$$

which indicates the relationship between time dummy and influencing factors. In equation (17), $z$ proved to be the logarithm of fractal dimension odds. The second component, a key part, is a logistic function

$$y=\frac{1}{1+e^{-z}}, \tag{18}$$

which indicates the squashing effect of fractal dimension growth. The third component is a ratio function, which represents the definition of output variable. Normalizing fractal dimension yields a probability measure, $p(t)$, which serves as a response variable, $y$, as follows

$$y=p(t)=\frac{D(t)}{D_{\max}}, \tag{19}$$

in which $D_{\max}$ can be let $d=2$ for simplicity. The relationships between logarithm of fractal dimension odds and time dummy can be expressed as

$$z=\ln(\frac{D(t)}{D_{\max}-D(t)})=\begin{cases}kt\\(kt)^2\end{cases}. \tag{20}$$

For the common logistic growth, we have $z=kt$, and for the quadratic logistic growth, we have $z=(kt)^2$. All analysis processes of logistic regression form a three-layer artificial neural network model, which is termed error back propagation (EPB) network (Figure 2).

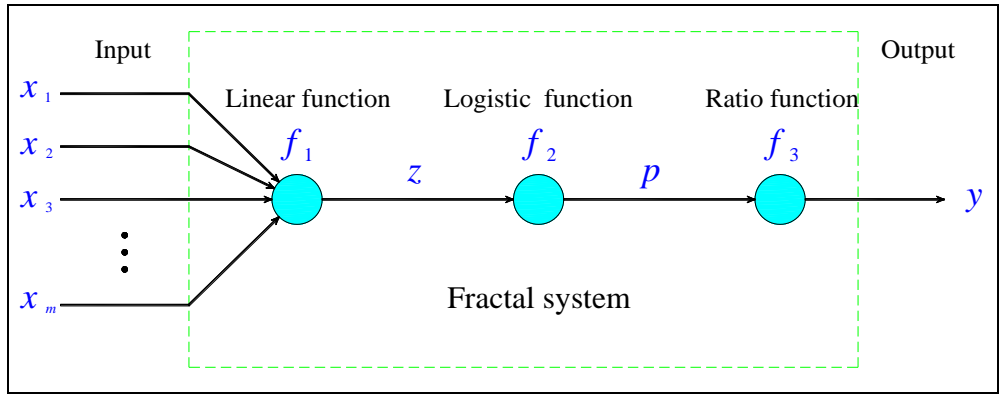

**Figure 2 A sketch map for three components of logistic modeling of fractal dimension curve of urban form and growth**

Note: The analytical framework of fractal-based on logistic regression comprises three components, which can be expressed as three function. The structure can be regarded as a three-layer EPB artificial neural network model.

The fractal dimension value of a complex system depends on measurement and calculation methods. So the method of fractal dimension determination influences the effect of logistic modeling of urban growth. There are at least five sets of methods to define fractal dimension (Takayasu, 1990). All these methods can be applied to different aspects of fractal cities. Among all these methods, two methods are commonly used to characterize urban morphology and growth. One is box-counting method (Benguigui *et al*, 2000; Feng and Chen, 2010; Jiang and Liu, 2012; Shen, 2002), and the other is radius-area scaling method (Batty and Longley, 1994; Frankhauser, 1998a; Jiang and Zhou, 2006; White and Engelen, 1993). The former is mathematically equivalent to grid method (Frankauser, 1998b), and the latter is equivalent to radius-density scaling method and can be replaced by radius-number scaling method (Batty and Longley, 1994; Longley *et al*, 1991). The fractal dimension defined by radius-area scaling is termed radial dimension, which is actually a local fractal parameter since it relies on the selection of measurement center. Comparatively speaking, the fractal dimension defined by box-counting method is a global dimension, which can also be termed grid dimension (Frankhasuer, 1998b). For a regular fractal, the box dimension may be equal to the radial dimension (Batty and Longley, 1994). However, for a random fractal, the two types of dimension values are not the same. Empirical studies suggests that the fractal dimension sequence measured using the box method can better exhibit S-shaped curve features. The box method can be divided into two measurement methods: one is fixed box method (Batty and Longley, 1994; Jiang and Liu, 2012; Shen, 2002), and the other variable box method (Benguigui *et al*, 2000; Feng and Chen, 2010). For the first method, use the same largest box for different years; for the second method, the largest box is determined by the city size in different years. The fractal dimension sequence based on fixed box method can better reflect the logistic process of space-filling in an urban administrative district, while the fractal dimension sequence based on variable box method can better reflect the logistic process of space filling in urbanized area. It can be seen that logistic regression modeling needs to be carried out according to specific research objectives.

It is necessary to compare fractal logistic regression based on fractal theory with logistic regression based on statistics. In multivariate statistical analysis, there is a method called logistic regression, including binary logistic regression and multinomial logistic regression. The multinomial logistic regression can be decomposed into binary logistic regression. There is an analogy between the binary logistic regression in multivariate statistical analysis and the logistic regression modeling for fractal dimension curves. Drawing a comparison between the conventional logistic regression and fractal-based logistic regression is helpful for understanding the principle and methods develop in this work. The similarities and differences between the two types of logistic regression are tabulated as follows (Table 1). Both the conventional logistic regression and fractal-based logistic regression bear an analogy with the above-mentioned three-layer EPB artificial neural network model.

**Table 1 A comparison between conventional logistic regression analysis and fractal-based logistic regression analysis**

| Type | Logistic regression in multivariable statistics | Logistic regression for fractal dimension analysis |
|---|---|---|
| **Rationale** | Logistic regression modeling | Logistic regression modeling |
| **Functions** | Three functions: step function, logistic function, linear function | Three function: ratio function, logistic function, linear function |
| **Input variables** | Three types: metric variable, rank variable, categorical variable | Three types: metric variable, rank variable, categorical variable |
| **Output variable** | Categorical variable | Metric variable |
| **Algorithm** | Maximum likelihood estimate (MLE) method | Ordinary least square (OLS) method |
| **Scope of application** | Statistical analysis: discrete selection analysis, categorical analysis, discriminant analysis, and so on | Fractal studies: explaining fractal growth, stage division of fractal dimension curve, and so on |

The above modeling and analysis methods can be extended to other branches of geography. Similar logistic modeling analysis can be conducted as long as there is logistic growth or generalized logistic growth phenomena. Therefore, the methods can be generalized to analyze urbanization curve, traffic network development, technology innovation diffusion, and so on. Urban form and growth belong to intraurban geography (De Keersmaecker et al, 2003), while transport network more belong to interurban geography (De Blij and Muller, 1997). Both intraurban geography and interurban geography involve urbanization. Urbanization is a process of urban-rural population replacement (Rao et al, 1988; Rao et al, 1989). There is analogy between fractal dimension curve and urbanization curve. The increase of urbanization level takes on squashing effect, and urbanization curves can be modeled by using logistic function or quadratic logistic function (Cadwallader, 1996; Davis, 1969; Pacione M (2009; Zhou, 1995). Define the level of urbanization as follows (Karmeshu, 1988; United Nations, 1980)

$$L(t) = \frac{u(t)}{u(t) + r(t)}, \qquad (21)$$

where $L(t)$ denotes the level of urbanization of time $t$, $u(t)$ and $r(t)$ represents urban population and rural population, respectively. Thus we have a logistic regression model as follows

$$\ln \frac{L(t)}{1 - L(t)} = \ln V(t) = b_0 + \sum_{j=1}^{m} b_j x_j(t), \qquad (22)$$

in which $V(t) = L(t)/(1-L(t)) = u(t)/r(t)$ refers to urban-rural ratio, representing another measure of urbanization (United Nations, 2004). Other notation is the same as equation (6). Using equation (22) to make logistic regression analysis, we can reveal the influencing factors of urbanization and distinguish between primary and secondary factors.

There is an inherent relationship between the development of transportation networks and the level of urbanization. In a similar way, the modeling methods can also be generalized to the growth curve of $\beta$ index of transport network. A network is composed of nodes and edges. The $\beta$ index is defined as the ratio of edge number $u$ to node number $v$, that is $\beta = u/v$. The largest number of edges is $u = v(v-1)/2$. Thus the maximum value of index is $\beta_{max} = (v-1)/2$. The growth curve of $\beta$ index can be modeled with Boltzmann equation or quadratic Boltzmann equation. Based on normalized variables, Boltzmann equation and quadratic Boltzmann equation change to logistic function and quadratic logistic function, respectively. Thus, a logistic regression model can be given as below

$$\ln \frac{\beta(t)}{\beta_{\max} - \beta(t)} = b_0 + \sum_{j=1}^{m} b_j x_j(t). \tag{23}$$

By using equation (23) for logistic regression analysis, we can reveal the influencing factors on the development of transport networks and then rank them in order of priority.

Further, this method can be extended to the study of the diffusion process of technological innovation. In a region, cumulative acceptance an innovation in time takes on an S-liked curve (Morrill et al, 1988). If the sigmoid curve can be described with a logistic function, it indicates a replacement process. The process of technological innovation diffusion is actually a process of replacing old and new technologies (Hermann and Montroll, 1972; Fisher and Pry, 1971). Therefore, logistic regression modeling can be used to find the influence factors of technology innovation diffusion. The model is as follows

$$\ln \frac{\varphi(t)}{1 - \varphi(t)} = b_0 + \sum_{j=1}^{m} b_j x_j(t). \tag{24}$$

where $\varphi(t)$ refers to the ratio of new technology to all technologies.

The novelty of this article lies in the invention of an urban growth analysis model based on fractal dimension and multiple explanatory variables. Similar studies seem to have not been reported before. The shortcomings of this study include three aspects. Firstly, the explanatory variables for urban growth in China are all statistical data. Compared with census data and big data generated from bottom to top, the confidence of statistical data is low. However, as an example of an analysis method, the problem is not significant. Secondly, there are no analysis cases of Western cities. The fractal dimension curve of urban growth in Europe and America meets a conventional logistic function, while the fractal dimension curve of urban growth in China satisfies a quadratic logistic function (Chen, 2018). Comparing the two modeling results between European and American cities and Chinese cities is more enlightening. Unfortunately, there is no system data for European and American cities. Thirdly, only the linear relationship between time dummy and influencing variables are taken into account. The variable relationships may be nonlinear, and the logistic regression may be replaced by other type of nonlinear models. All these problems remain to be explored in future.

# 4 Conclusions

So far, the framework construction of the logistic regression modeling method for the fractal dimension curve of urban form and growth has been completed preliminarily. The main points of this study can be summarized as follows. First, based on the sigmoid functions of fractal dimension curves, a logistic regression modeling method can be developed for multivariate analysis of urban growth. The dependent variable is the logarithm of fractal dimension odds, and the covariates include varied possible factors which influence city development. By making stepwise regression analysis, we can determine the significant factors affecting urban growth and rank them in order of importance. The statistical testing method is readily available and can be judged directly using the statistics of multiple regression analysis. Second, logistic regression analysis of fractal dimension of urban form and growth can be divided into longitudinal analysis and transverse analysis. By means of time series of fractal dimension of a city, we can make a longitudinal logistic regression analysis for urban growth. This type of study belongs to intraurban geography. By using cross-sectional dataset of an urban system, we can make a transverse logistic regression analysis. This type of study belongs to interurban geography. For a system of cities, these two methods can complement each other. Third, the method of modeling can be generalized to other phenomena of logistic growth. The level of urbanization can be modeled by using logistic function or quadratic logistic function. Therefore, the modeling method can be applied to urbanization curve. The $\beta$ index of transport network can be described with Boltzmann's equation or quadratic Boltzmann's equation. Based on normalized variable, Boltzmann's equation and quadratic Boltzmann's equation can be turned into logistic function and quadratic logistic function, respectively. So, the logistic modeling method can be applied to the $\beta$ index of transport network. Moreover, technology innovation diffusion is a type of replacement dynamics, the increase in the proportion of new technologies is manifested in an S-shaped curve and can be analyzed by similar method of logistic modeling.

**Acknowledgement:**

This research was sponsored by the National Natural Science Foundation of China (Grant No. 42171192). The support is gratefully acknowledged.